\documentclass[prl, preprint,showkeys, amsmath,amssymb, aps, ]{revtex4-2}

\UseRawInputEncoding
\usepackage{lineno}
\usepackage{xcolor}
\usepackage{siunitx}
\usepackage{graphicx}
\usepackage{dcolumn}
\usepackage{bm}

\linenumbers\relax 
\nolinenumbers

\begin{document}

\preprint{APS/123-QED}

\title{\textbf{
Coherent shear wave in crystalline Si upon sub-threshold femtosecond laser-induced excitation
} 
}%

\author{Angel Rodr\'{i}guez-Fern\'{a}ndez}
 \email{Contact author: angel.rodriguezfernandez@diamond.ac.uk}
\author{Jan-Etienne Pudell} 
\author{Roman Shayduk}
\author{James Wrigley}
\author{Alejandro Fraile-Gimeno}
\author{Wonhyuk Jo}
\author{Johannes M\"{o}ller}
\author{Alexey Zozulya}
\author{J\"{o}rg Hallmann}
\author{Anders Madsen}
\affiliation{European X-ray Free-Electron Laser Facility, Holzkoppel 4, Schenefeld, DE-22869}

\author{Pablo Villanueva-Perez}
\affiliation{ Division of Synchrotron Radiation Research and NanoLund, Department of Physics, Lund University, Lund, SE-22100 
}

\author{Zdenek Matej}
\affiliation{MAX IV Laboratory, Lund University, Lund, SE-22100}

\author{Thies J. Albert}
\author{Dominik Kaczmarek}
\author{Klaus Sokolowski-Tinten}
\affiliation{Department of Physics, Universität Duisburg-Essen, Lotharstr. 1,  Duisburg, DE-47057.}
\affiliation{Center for Nanointegration Duisburg-Essen, Universit\"{a}t Duisburg-Essen, Carl-Benz-Str. 199, Duisburg, DE-47057}

\author{Antonowicz Jerzy}
\affiliation{Faculty of Physics, Warsaw University of Technology, Koszykowa 75, Warsaw, PL-00662}

\author{Oleksii I. Liubchenko}
\author{Houri Rahimi Mosafer}
\author{Ryszard Sobierajski}
\affiliation{Institute  of Physics of the Polish Academy of Sciences, Aleja Lotników 32/46,  Warsaw, PL-02668}

\author{Javier Solis}
\author{Jan Siegel}
 \email{Contact author: j.siegel@io.cfmac.csic.es}
\affiliation{ Laser Processing Group, Instituto de Optica (IO-CSIC), Consejo Superior de Investigaciones Cient\'{i}ficas, CSIC, , Madrid, E-28006 }



\date{\today}

\begin{abstract}
Femtosecond laser processing of semiconductor wafers is driven by applications in the electronics and photonics industry. 
X-ray free-electron lasers are powerful probes for revealing the ultrafast dynamics of the induced changes. 
We present a novel technique that provides time and depth resolved snapshots of the strain field upon laser excitation of bulk crystalline Si. 
At fluences below the melting threshold, coherent shear waves 
were found to propagate from the laser-excited surface into the bulk.
Simulations support a propagation speed of \SI{5.8}{\kilo \meter \per \second}, slower than the longitudinal speed of sound, \SI{8.4}{\kilo \meter \per \second}.
\end{abstract}

\maketitle


Over the last decades, the countless potential applications in technology and industry have further enhanced the strong interest in the use of ultrafast lasers for material processing \cite{Stoian23}.
One key aspect is the reduced thermal load with respect to nanosecond pulse and continuous wave laser processing, enabling the fabrication of smaller and sharper features, even below the diffraction limit \cite{Chambonneau21,Stoian20}. 
By using laser pulses that are shorter than the time it takes for the strongly excited electron subsystem to transfer its energy to the lattice (typically a few picoseconds), highly non-equilibrium states can be accessed and structural changes triggered within a few hundred femtoseconds \cite{Zier15}. 
Laser structuring bulk semiconductors with ultimate precision requires an understanding of the complex processes involved.
The different transient states and their dynamics can be investigated by optical pump-probe techniques, which are inherently surface sensitive and have been used to confirm non-thermal melting in semiconductors \cite{Casquero_2022,Sokolowski98}.
Also, strain or shock waves that propagate deep into the material \cite{Hu10} at speeds of up to \SI{10}{\kilo\meter\per\second} have been investigated with optical probe techniques \cite{Smith2012} and the interference of laser coherent acoustic pulses in transparent materials \cite{Gusev18, Wang2020}.
However, these are indirect techniques since they rely on monitoring the optical properties of materials, rather than the structural phase, which makes x-ray diffraction techniques indispensable.

Early works using time resolved x-ray diffraction (TR-XRD) were able to confirm laser-induced lattice distortions in semiconductors \cite{Siders99,Sokolowski01, Rousse_2001,Larson82,tischler_1988,Loveridge01,DeCamp05,Lings06,jo_2022}.
However, most of these studies suffered from limited temporal resolution 
and/or low x-ray flux.
In most of the works till today, the diffraction signal recorded by the detector was a depth-average over a few microns of the sample.
This limitation can be mitigated by studying thin film samples 
\cite{Sokolowski01,Lindenberg08,pudell_layer_2018,jo_2022}.
Yet, at least for the case of semiconductors, occupying a  dominant position in the electronics industry and silicon photonics \cite{Almeida04}, bulk materials (wafers) are commonly used in real-world processing applications. 
With the advent of x-ray free-electron lasers (XFELs), which provide spatially coherent, hard x-ray pulses with high flux and ultrashort duration \cite{LCLS10,EUXFEL20},  new powerful structure-sensitive techniques for studying the dynamics of ultrafast processes, such as strain, phonon oscillations or melting have emerged \cite{Trigo02,Alon15,Zalden19,Jo2021,Lee_16, pandolfi2022,Shayduk22,ANTONOWICZ2024,Marais23}.

Developing an experimental tool capable of measuring femtosecond laser-induced structural changes and strain fields occurring on time scales from sub-picoseconds up to nanoseconds with sub-picosecond and depth resolution is highly desired.
Several experimental strategies have been proposed to this end:
Bright and dark field x-ray microscopy have demonstrated capabilities for imaging the effect of laser-induced shock waves in the vicinity of crystal defects, such as dislocations \cite{Marais23,Irvine25,Carlsen22}. 
Near field x-ray diffraction microscopy
allows recording diffracted wavefronts  from single crystals \cite{ARF18}.
Rodriguez-Fernandez et al. used a variant of x-ray ptychography to analyze diffracted wavefronts distorted by surface strained crystals \cite{ARF21}.
To simulate the wavefronts, a formalism based on ultrafast dynamical diffraction (UDD) theory was used \cite{ARF18,ARF21,ARF23}.
The UDD signal is the constructive interference of all the x-ray beams diffracted along the depth of the crystal in an area known as the Borrmann fan (BF) (purple triangle in the sketch of Fig. 1) \cite{Borrmann49,Takagi62,Bat64,Authi01}. 
The spatial distribution of the maxima in the UDD, denoted as echoes, is the Fourier transform of the reflectivity curve of the crystal lattice at a particular time \cite{ARF21,ARF23}.
More information about the UDD is given in the End Matter.

In this work, we used near field x-ray diffraction microscopy to record single-shot images of the BF fine structure of a Si single crystal in Laue diffraction geometry at the European XFEL.
We present the laser-induced transient changes of the BF signal as a function of laser fluence and pump-probe time delay in an excitation regime below the melting threshold.
We retrieve time and depth information from the transient signal together with numerical calculations that combine UDD simulations with a 3D version of the model of Thomsen et al. \cite{Thomsen96} to describe the generation and propagation of picosecond strain pulses. 
Surprisingly, good agreement with the experimental data is only achieved by introducing 
a coherent shear wave
propagating at the transversal speed of sound, $\upsilon_\text{T}=\SI{5.8}{\kilo\meter\per\second}$ \cite{Smith2012} into the depth, i.e. slower than the longitudinal speed of sound, $\upsilon_\text{L}=\SI{8.4}{\kilo\meter\per\second}$. 
These observations challenge our current understanding of strain wave generation in laser-irradiated high-symmetric semiconductors, where, due to the quasi 1D excitation geometry, i.e.\ the laser spot size being much larger than the optical excitation depth, only longitudinal strain waves are expected.  

\begin{figure}[b] 
\includegraphics[scale = 0.22]{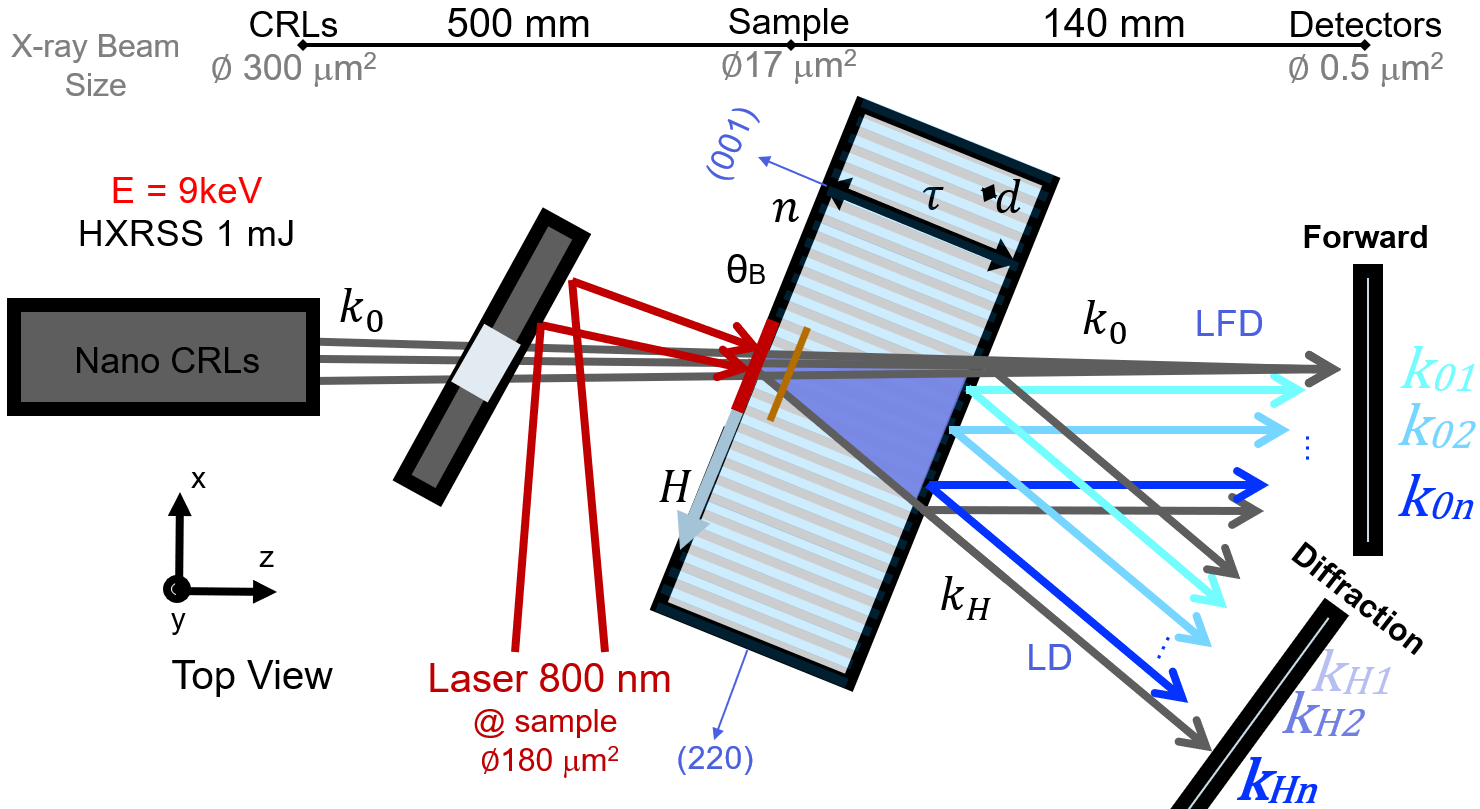}
\caption{\label{fig1:Sketch_MID} Scheme of the experimental setup. The Borrmann fan is denoted with a purple triangle. Sample and distances not to scale.
}
\end{figure}


The experiment was carried out at the Materials Imaging and Dynamics instrument \cite{Madsen21}.
The  SASE~2 beamline delivered \SI{10}{\hertz} self-seeded $\sigma$-polarized x-ray pulses at a photon energy of \SI{9}{\kilo\electronvolt} \cite{Liu23}.
Two Si $(220)$ channel cut monochromators were used to clean the spectrum.
A sketch of the experimental setup is presented in Figure~\ref{fig1:Sketch_MID}.
The x-rays were focused onto the detector plane to a spot size of $0.5\times \SI{0.5}{\micro\meter^{2}}$ (FWHM) using $10$ beryllium compound refractive lenses with \SI{50}{\micro\meter} radius of curvature, which provides a focal length of \SI{640}{\milli\meter}. 
Two x-ray microscopes were used to collect the diffraction and forward signal.
Each microscope composed of a \SI{20}{\micro\meter} thick Ce-GaGG crystal, a 20x lens and an Andor Zyla $5.5$ detector.
The x-ray microscope in diffraction was used to collect the UDD signal discussed in this manuscript.


The sample was a \SI{300}{\micro\meter} thick $<001>$ orientation Si wafer.
The crystal misorientation was determined to be less than \SI{2}{\milli rad}. 
The beam size at the sample position was $\varnothing$\SI{17}{\micro\meter}. 
The sample was aligned to the $(220)$ symmetric Laue diffraction condition; Fig. S1 in the supplementary material (SM) presents an experimental rocking scan \cite{SM20}.
In this geometry, the diffraction signal is only sensitive to variations of the lattice spacing projected onto the momentum transfer $H$, e.g.\ shear strain waves propagation along the surface normal, but not to distortions along the $[001]$ direction, e.g.\ longitudinal strain waves propagating normal to the surface.

An optical laser with a central wavelength of \SI{800}{\nano\meter}, a pulse duration of \SI{25}{\femto\second} and a maximum fluence of \SI{300}{\meter J \per c \meter^{2}} was operated in an on-and-off mode to excite the front surface of the sample \cite{Palmer19}.
The optical laser traveled almost collinear with the x-ray beam, using an in-coupling mirror with a \SI{3}{\milli\meter} hole to allow the x-ray beam to be transmitted.
The laser passed a circular aperture of \SI{4}{\milli\meter} before it was focused on the sample, using a lens, to a near circular spot with a measured waist ($1/e^2$ radius) $w$ = \SI{100}{\micro\meter}.
In view of the large laser spot size with respect to the thickness of the layer within which the laser energy is deposited (a few hundreds of nanometers), we can assume that the strain wave propagates perpendicular to the surface.
The fluence (F) of the laser was controlled using a rotatable half-wave plate in combination with a polarizing beam splitter \cite{Shayduk22}.
Beam waist and fluence calibration were performed using single pulse irradiations above the material modification threshold, as presented in Fig.~S2 \cite{SM20}. 
These measurements yielded the melting threshold of the Si ($F_\text{m} = \SI{112}{\meter J \per c\meter^{2}}$) using the methods in \cite{Liu82,Bonse2006,Lechuga23}. 
A camera located after a mirror in the laser path was used to monitor laser beam pointing and intensity variations.
An in-situ microscope constantly monitored the absence of any damage induced by either pump or probe pulses on the sample surface.

\begin{figure}[b] 
\includegraphics[scale = 0.20]{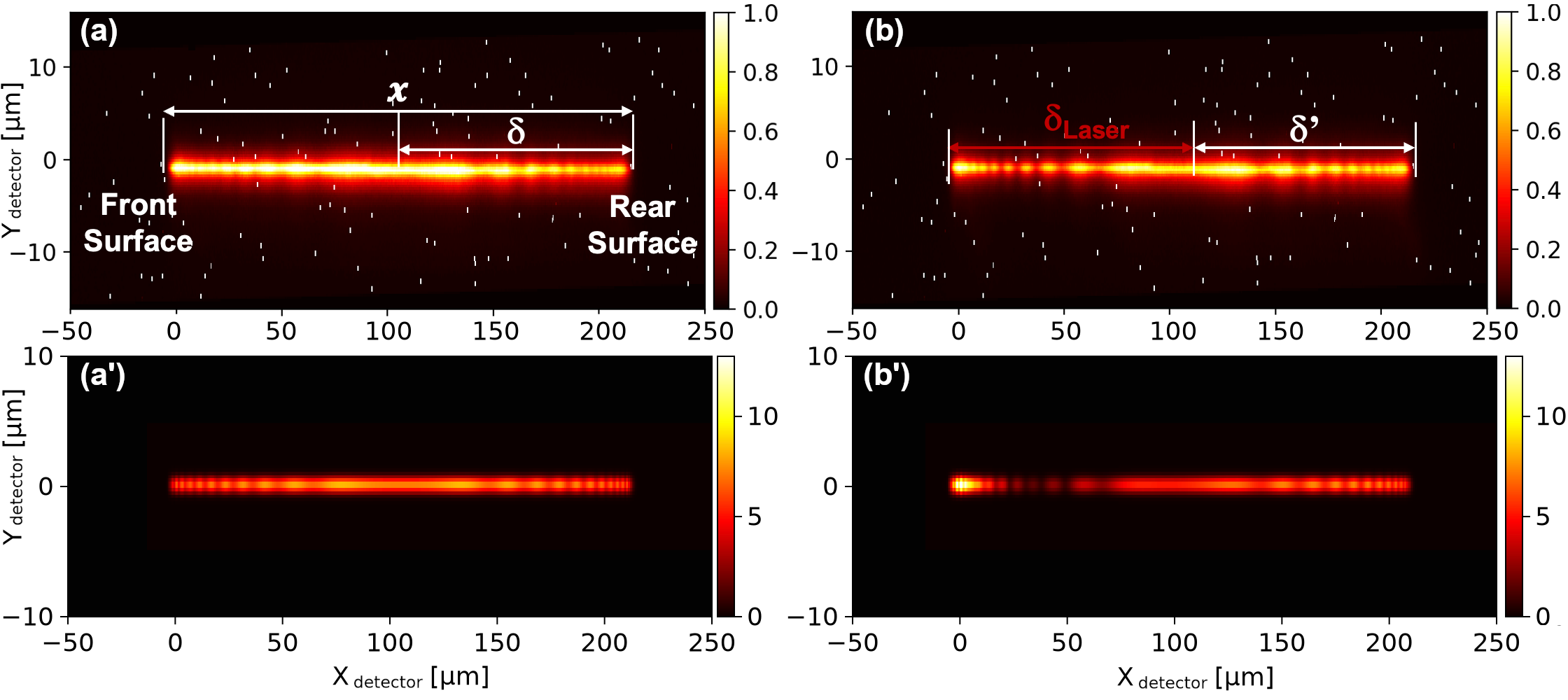}%
\caption{\label{fig2:Expvssim_2D} 
Single pulse x-ray diffraction wavefront of a Si \SI{300}{\micro\meter} thick crystal for the symmetric Laue $220$ reflection at \SI{9}{\kilo\electronvolt} in horizontal geometry. (a) No laser and (b) \SI{900}{\pico\second} after a single pulse from a \SI{800}{\nano\meter} femtosecond laser with $F = \SI{50}{m J \per c m ^{2}}$. Simulated diffraction wavefronts for the same experimental parameters, (a') No strain and (b') using the strain model describe in End Matter.
Here, $\delta$ is half of the undistorted signal length along the $x$ direction and $\delta_\text{laser}$ and $\delta'$ are the distorted and undistorted signal for the pumped wavefront, respectively.
}
\end{figure}


Figure \ref{fig2:Expvssim_2D}(a) shows the diffraction signal from the Si $(220)$ reflection recorded 
in absence of laser excitation.
The pattern corresponds to the BF and the intensity modulation is a result of the UDD.
Figure S3 in \cite{SM20}, shows the horizontal profiles of the diffraction patterns. 
For comparison, Fig. \ref{fig2:Expvssim_2D}(b) shows the pattern recorded upon laser excitation at a $F = \SI{50}{m J\per c m^{2}}$ and a pump-probe delay of $t = \SI{900}{\pico\second}$, revealing large signal changes at a fluence well below the melting threshold.
While the intensity distribution without pump laser in Fig.~\ref{fig2:Expvssim_2D}(a) is essentially symmetric with respect to the center of the wave field, the pumped signal in Fig.~\ref{fig2:Expvssim_2D}(b) shows a strong decrease of the BF signal originated from the front surface.

Due to the interference nature of the UDD mechanisms, a quantitative interpretation of the static and transient signals requires modelling.
To this end, the UDD code used in \cite{ARF18, ARF21} was seeded with an analytical laser-induced strain model based on the work by Thomsen et al. \cite{Thomsen96}.
This model assumes a component perpendicular to the surface of the crystal (longitudinal direction) for which a bipolar strain wave propagates at $\upsilon_\text{L}$ and a second component parallel to the surface, $(hh0)$, and orthogonal to the wave propagation direction, in which a uni-polar transverse deformation propagates into the crystal with the transversal speed of sound, $\upsilon_\text{T}$.
An amplitude factor, $I_{adj}$, scales the strain deformation between the longitudinal and orthogonal components along the depth.
The best match to the experimental data happens for $I_{adj} = 1$.
More information about the simulations is presented in the End Matter.
Fig.~\ref{fig2:Expvssim_2D}(a') and (b') presents the simulations performed with the UDD code for the same experimental conditions as in Fig.~\ref{fig2:Expvssim_2D}(a) and (b).
In the pristine case shown in Fig. \ref{fig2:Expvssim_2D}(a'), the simulated signal matches well the one observed in experiment Fig.~\ref{fig2:Expvssim_2D}(a).
For the excited sample Fig.~\ref{fig2:Expvssim_2D}(b'), the simulated signal exhibits a similar depression as observed in the experimental data.
It should be emphasized that the calculated signal of the pumped scenario strongly depends on the laser absorption depth ($\zeta$) and fluence. 
Therefore, we treat the $\zeta$ as a parameter to account for non-linear absorption effects (i.e.\ free-carrier and multi-photon absorption) in silicon. 
For $F = \SI{50}{m J \per c m ^{2}}$ an effective $\zeta$ of \SI{300}{\nano\meter} gave the best match to the experimental data, as demonstrated in the Fig.~S4 \cite{SM20}.

\begin{figure}[b] 
\includegraphics[scale = 0.40]{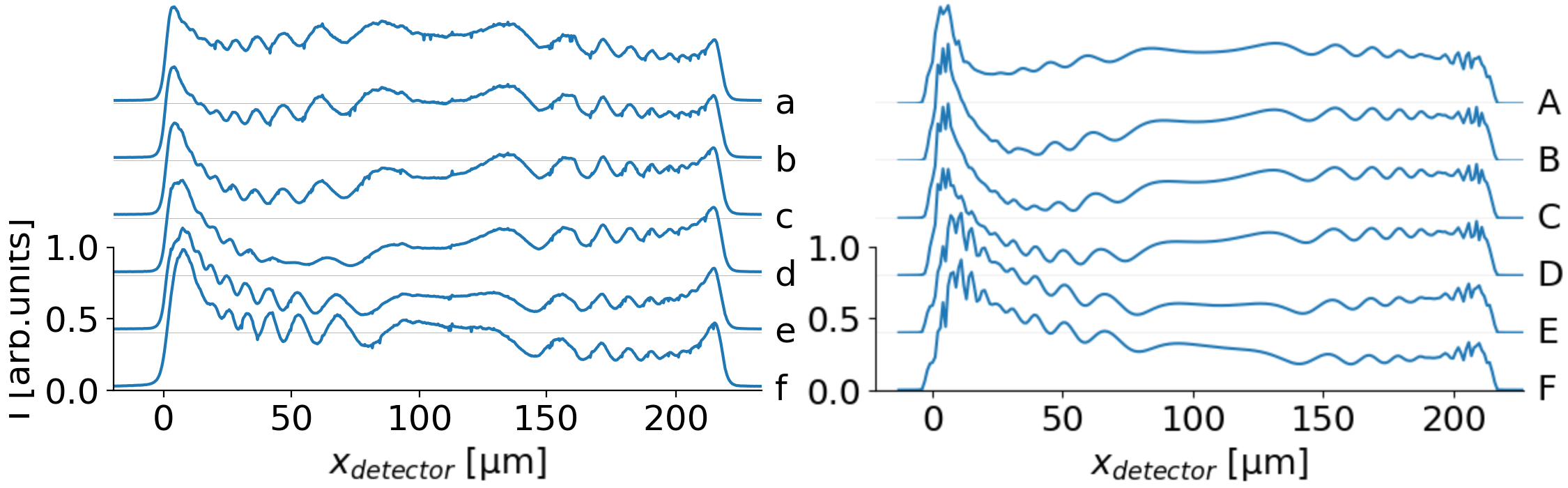}%
\caption{\label{fig3:Fluence_scan}  
(Left) Experimental diffracted x-ray signal profiles after pump laser excitation at an x-ray delay of \SI{900}{\pico\second} for $F=$ a: \SI{20}{mJ\per cm^{2}}, b: \SI{40}{mJ\per cm^{2}}, c: \SI{50}{mJ\per cm^{2}}, d: \SI{70}{mJ\per cm^{2}}, e: \SI{85}{mJ \per cm^{2}} and f: \SI{100}{mJ\per cm^{2}}. (Right) Corresponding simulations for the same experimental conditions using two different $\zeta$; \SI{300}{\nano\meter} for A, B, C and D and \SI{100}{\nano\meter} for E and F.
}
\end{figure}

We have investigated the influence of the laser fluence on the lattice distortions, at constant delay $t = \SI{900}{\pico\second}$.
The fluence range explored was from \SI{20}{m J \per c m ^{2}} to \SI{100}{m J \per c m ^{2}}.
The experimental profiles are shown in Fig.~\ref{fig3:Fluence_scan} (left).
A progressive decrease of the x-ray diffraction signal amplitude as a function of the laser fluence is observed.
The dip in the amplitude related to the laser excitation is located in all cases around \SI{50}{\micro\meter}, except for $F=$ \SI{100}{m J \per c m ^{2}}, close to ($F_\text{m}$).
For this fluence, as is observed in the lower profile of Fig.~\ref{fig3:Fluence_scan}(f), there is a change in the slope of the signal profile, which could be related to a change in the effective $\zeta$ due to the higher fluence.
We have simulated the corresponding profiles using our strain model, the results being presented in Fig.~\ref{fig3:Fluence_scan} (right). 
Overall, there is a progressive intensity reduction around the position \SI{50}{\micro \meter} of the signal, as observed experimentally.
However, the simulations shown in Fig.~\ref{fig3:Fluence_scan}; A and B predict a stronger suppression than observed experimentally. 
Looking in more detail, it can be seen that A better matches the experimental result b. 
A similar behavior is shown in Fig.~S5 \cite{SM20}, where we present the simulations for \SI{10}{m J \per c m ^{2}}, which shows a more satisfactory match to Fig.~\ref{fig3:Fluence_scan} a.
This behavior is indicative of a reduced energy deposition at low fluences, which is consistent with the presence of a non-linear absorption mechanism.
For moderate fluences C and D, the results show a good agreement with the experimental data c and d, respectively.
For Fig.~\ref{fig3:Fluence_scan} E and F, the good match observed and the change in the slope at the detector center was only possible to obtain by reducing the effective $\zeta$ to \SI{100}{\nano \meter}, in agreement with the expected fluence-dependent effective $\zeta$. 
Simulations performed with a single $\zeta$ are shown in Fig.S6 (SM) \cite{SM20}, showing a worse match at higher fluences.

\begin{figure}[b] 
\includegraphics[scale = 0.45]{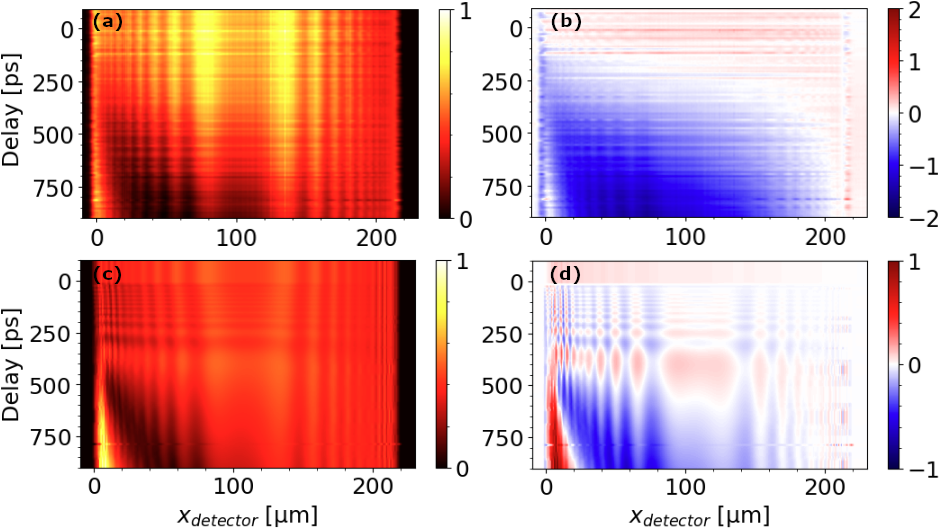}%
\caption{\label{fig4:Delayscan} 
Waterfall plots (a) of the experimental diffraction signal upon laser excitation at $F=\SI{50}{mJ \per cm^2}$ as a function of pump-probe delay time and the X detector coordinate.
(b) Difference between experimental pumped, as shown in (a), and un-pumped experimental traces.
(c) Waterfall plot of the simulated profiles for the delay scan range using F =\SI{50}{mJ \per cm^2}, $\zeta$ = \SI{300}{\nano \meter} and $\upsilon_\text{T}$.
(d) Difference between simulated pump and un-pump signal as presented in (c).
}
\end{figure}

Besides the laser fluence, the delay time between the NIR laser pump and x-ray probe beams is expected to strongly affect the recorded signal and to allow unraveling the strain wave propagation dynamics.
A waterfall representation of the experimental profiles as a function of X-detector and pump-probe time delay (\SI{-100}{\pico\second} to \SI{900}{\pico\second}) is presented in Fig.~\ref{fig4:Delayscan}(a).
The experimental data features the above-mentioned near-surface signal depression whose amplitude increases with time and whose position moves along the crystal depth (most pronounced in the range from \SI{300}{\pico\second} to \SI{900}{\pico\second}). The same behavior was observed in a \SI{100}{\micro \meter} Si wafer as presented in Fig.~S7 \cite{SM20}.
Figure~\ref{fig4:Delayscan}(b) presents the calculated difference between the experimental pumped and un-pumped traces, in order to highlight the relative change by suppressing the oscillations.
Qualitatively, the same trend is observed in the simulated profiles presented in the corresponding waterfall plots in Fig.~\ref{fig4:Delayscan}(c,d) using $\upsilon_\text{T}$ in the model described above. 
This includes the strong signal arriving from the front surface predicted at longer delays, forming a narrow yellow triangle near the surface.
For comparison, the calculations performed for $\upsilon_\text{L}$ yield a much wider yellow triangle (displayed in the SM Fig.~S8 \cite{SM20}.), inconsistent with the experimental data.
Moreover, the reduction of the intensity in the calculations for $\upsilon_\text{L}$ appears to start around \SI{200}{\pico \second} after the laser excitation, earlier than observed in the experimental data.
Selected profiles at different delay times between pump and probe for the experimental and simulated signals for the case of $\upsilon_\text{T}$ are presented in the SM Fig.~S9 \cite{SM20}.
In Fig.~S10 from \cite{SM20}, we present the diffraction profiles for longer delays collected during the experiment. 
Fig.~S11 in \cite{SM20} shows the forward diffraction profiles for several short delays.

The overall good match of the simulations with the experimental data, as shown in ~Fig.~\ref{fig4:Delayscan}, suggests that femtosecond laser excitation of Si -- a cubic material -- leads to the generation of a shear strain wave with an amplitude comparable to the expected bi-polar longitudinal wave. 
This experimental evidence of a shear wave under the present conditions is a surprising result because such large amplitude shear waves have never been considered before at low fluences for cubic materials, only at much stronger excitation, where the material deforms plastically upon rapid shock compression \cite{Smith2012,pandolfi2022,Lee_16}.
Two principal possibilities for the generation of
shear waves in crystals have been considered: (1) Breaking the axial symmetry by choosing either off-axis geometries and acoustically anisotropic media \cite{Pezeril_07,Matsuda_04,Lemke_18}, or piezo-electric/electro-strictive materials \cite{Lejman_14,Wen_09,Juve_20, Gu_23};
(2) Using small laser spot sizes ($\approx$\SI{1}{\micro\meter}), so that the lateral stress gradients become comparable to the pressure variation along the surface normal leading to acoustic diffraction and mode conversion \cite{Rossignol05,VERES2013}. 
While Si is acoustically anisotropic and electro-strictive, possibility (1) seems unlikely since we have used a silicon wafer cut along high symmetry directions with minimal miscut.
Electro-striction can create high (anisotropic) stress given the high laser intensity in the \SI{}{\tera \watt \per c m ^{2}} range, but it is very short-lived (i.e. \SI{25}{\femto \second} pulse duration).
Therefore, in laser-excited Si isotropic electronic (deformation potential coupling) and thermo-elastic stress dominate the acoustic response. 
To check (2), we have compared COMSOL-simulations of the acoustic response after laser excitation for a ''small'' (\SI{2}{\micro\meter} FWHM) and a ''large'' (\SI{100}{\micro \meter} FWHM) laser excitation spot on the surface (see Fig.~S12).
While our small focus simulations confirm the results in \cite{Rossignol05,VERES2013}, where acoustic diffraction and mode conversion play an important role, these effects are very much suppressed for the large focus, in particular, for the pump-probe time delays considered in this study ($<$ \SI{1}{\nano \second}) and in the small central area probed by the x-rays.
Molecular dynamics simulations may be key to understand the observed process \cite{Djafar2022,ROUSSEAU2025,Koziol2026}.

In previous studies of single crystals using TR-XRD in both Bragg and Laue geometry, Loveridge-Smith and collaborators used a nanosecond laser to produce longitudinal shock waves propagating inside a Si crystal \cite{Loveridge01}.
When a clear distorted signal was observed in the Bragg geometry, no clear distortion was observed in the Laue geometry. 
On this basis, the authors state that no deformation was observed in the orthogonal direction to the propagation of the longitudinal shock wave. 
It is important to remember that in the present work, we are resolving the fine structure inside the diffraction peaks and it is here that we see the variations in intensity and position of the maxima.
DeCamp et al. used TR-XRD to study Ge and InSb in Laue and Bragg geometry. 
The changes in diffraction signal observed along the rocking curves in Laue geometry were explained with a possible miscut of the sample of 3 mrad. 
From the simulations performed in our work, the miscut values needed to show signals as high as those observed experimentally in the absence of a shear wave are well above \SI{8}{\milli rad} (as presented in Fig.~S13 \cite{SM20}). 
Lings et al. used TR-XRD in the asymmetric Ge $(20\overline{2})$ Laue geometry to study the effect of nanosecond laser excitation \cite{Lings06}. 
In our opinion, the inclusion of the orthogonal component could have improved the resemble between simulations and experiment.
Lee and co-workers developed a Bragg TR-XRD technique that allowed using asymmetric reflections to follow a combination of both the longitudinal and shear deformations of a cubic lattice up to the depths related to the Bragg extinction length after femtosecond laser excitation \cite{Lee_16}.

To conclude, we have investigated the fine structure of the diffraction signal from the BF for a \SI{300}{\micro\meter} thick Si wafer for the $(220)$ reflection in symmetric Laue geometry. 
Variations of the echoes inside the BF upon femtosecond NIR laser excitation at the front surface as a function of delay and laser fluence have been recorded in a single pulse pump-probe scheme.
The experimental data exhibit large changes of the echoes demonstrating high sensitivity to transient lattice distortions in the crystal caused by shear strain wave propagating perpendicular to the surface into the depth of the material.
Already at a delay of \SI{300}{ps}, we observe a clear response in the diffracted signal. 
Moreover, simulations based on the Thomsen analytical model \cite{Thomsen96} resemble the experimental data for short delays and fluences below the melting threshold.
The model fits reasonably well the experimental data collected at $F = \SI{50}{ m J \per c m ^{2}}$, using an effective $\zeta$ of \SI{300}{\nano\meter} as a single fit parameter. 
From the simulations, we obtained a propagation speed for the orthogonal deformation close to $\upsilon_\text{T}$.

Our results challenge previous x-ray work using Laue geometry diffraction \cite{Loveridge01,Lings06}, where no orthogonal deformation was observed after laser excitation.
We attribute the difference with respect to the previous x-ray studies to our capability of resolving the fine structure in the diffraction peak.
Interestingly, the presence of the orthogonal distortion agrees with work performed at higher fluences \cite{Smith2012, pandolfi2022}.
Our results could be key to understand the starting mechanism of the phase transformation from diamond Si-I to Si-II phase as discussed in \cite{pandolfi2022}, in which a longitudinal elastic compression could trigger an expansion in the orthogonal direction. 

The study of an asymmetric Laue reflection might resolve simultaneously the two possible waves propagating along the longitudinal direction. 
Due to the high sensitivity of the UDD signal to the lattice distortions, this method could help to unveil ultrafast processes at higher fluences, such as ultrafast melting, ablation and shock wave generation, present upon femtosecond laser processing of single crystal. 





\begin{acknowledgments}
We acknowledge European XFEL in Schenefeld, Germany, for provision of x-ray free-electron laser beamtime at MID and would like to thank the staff for their assistance. 
Data recorded for the experiment at the European XFEL are available at doi: 10.22003/XFEL.EU-DATA-004977-00.
This research was supported in part through the Maxwell computational resources operated at Deutsches Elektronen-Synchrotron DESY, Hamburg, Germany.
This work was partly funded by the Consejer\'{i}a de Educaci\'{o}n, Ciencia y Universidades (Comunidad de Madrid, (Spain)) through the MATRIX-CM project (TEC-2024/TEC-85) and by MCIN/AEI/10.13039/501100011033 through HyperSpec grant (PID2023-148178OB-C22).
This work was supported by  ERC-2020-STG, 3M-FLASH (948426).
DK, TJA, and KST acknowledge by the Deutsche Forschungsgemeinschaft (DFG, German Research Foundation) through Project 638 No.~278162697-SFB 1242.
The access to the European XFEL was supported by a grant of the Polish Ministry of Science and Higher Education- decision no.~2022/WK/13.
This work was supported by the National Science Centre, Poland, grant agreement No 2021/43/B/ST5/02480.
We would like to thank J. Domagala for the support in the pre-characterization of the samples.
We would like to thank G. Carbone, U. Staub, A. Diaz, L. Horak, L. Samoyloba and I. Petrov., K. Appel and P. Zalden for the fruitful scientific discussion.
ARF would like to thank K. Finkelstein for many hours of discussion about dynamical diffraction and showing the first steps in the world of pump-probe in single crystals.
\end{acknowledgments}


A.R.F and J.Si conceptualized the work; R.Sy. and O.I.L. pre-characterized the samples; A.R.F., J.Si, J.E.P., and J.H. planned the experiment; A.R.F., J.E.P., R.Sh., W.J., J.M., A.Z., J.H, P.V.P., Z.M., T.J.A., D.K., K.S.T., A.J., O.I.L., H.R.M., R.Sy., J.So., and J.Si performed the experiment; A.R.F. analyzed the data with contributions of J.W., P.V.P., Z.M., T.J.A. and A.F.G.; A.R.F., and J.W. performed the simulations based in UDD; R.Sh.performed the COMSOL simulations; J.Si., K.S.T. and A.R.F. wrote the manuscript with the contribution and discussion of all authors.

\section{End Matter}

\subsection{About the x-ray Echoes}

Ultrafast dynamical diffraction (UDD) is a process in which multiple diffracted beams, denoted as echoes, are generated at the exit surface of a crystal, both in the diffraction and forward directions \cite{Bushu08,Shvy12,Shvy13}.
Only the x-rays that arrive at the surface of the crystal with the same phase will contribute to the signal. 
Due to the different paths of the x-ray photons in the crystal, each of these interferences exits the surface at a different time, which results in a delay with respect to the next x-ray echo of a few femtoseconds. 
The UDD process as presented in this experimental geometry is a two wave interference problem where only one crystal diffraction plane is excited. 
Due to this, all photons in the echoes share the same mono-chromaticity and divergence properties, and can be defined as delayed copies of the same x-ray pulse; thus the term “echoes” in resemblance to sound echoes. 
The distribution of these photon beams at the exit surface of the crystal in real space relates to the reflectivity for a particular moment in time of the crystal lattice along the depth. 
The entire UDD process occurs in the time that the x-ray photons need to travel through the crystal at the speed of light, e.g.  \SI{1}{\pico\second} for a \SI{300}{\micro\meter}  thick crystal. 
In this way, the measured spatial distribution of the echoes corresponds to a snapshot of the crystal lattice depth profile at the time in which the diffraction was collected.

\subsection{\label{app:trigonometric relation} Trigonometric approximation to understand the Borrmann fan distortion at short delays}

\begin{figure}[b] 
\includegraphics[scale = 0.15]{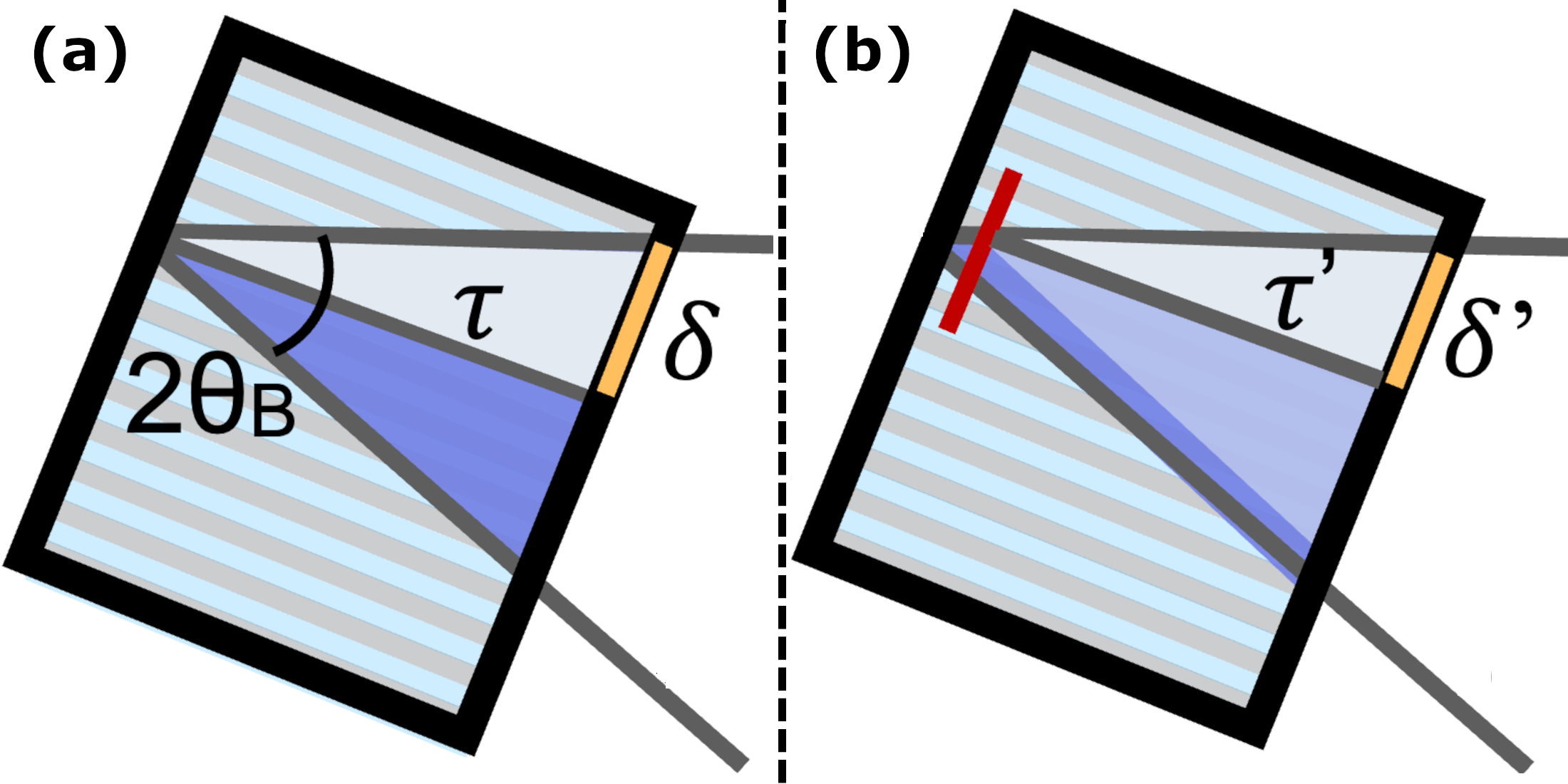}
\caption{\label{EM1:Sketch_BorrmannFan} Sketch of the BF for (a) pristine and (b) distorted crystal cases.
In (b) a red line represents the strain wave propagating along the crystal. 
A white triangle is painted to denote the area from the BF that is not distorted.
}
\end{figure}
As sketched in Fig.~\ref{EM1:Sketch_BorrmannFan}(a), for a pristine sample the elongation of the signal at the exit surface, $x$, is equivalent to two times the cathetus, $\delta$, of the right-angle triangle defined by the beam trajectory and the thickness of the crystal, $\tau$. 
Here, $\delta$ is related to the thickness $\tau$ and can be converted using $\delta = \tau \tan(\theta)$, where $\theta$ represents the diffraction angle.
For a crystal with $\tau = \SI{300}{\micro\meter}$ diffracting at an angle $\theta$ of \SI{21.021}{\degree},  $\delta$ will be of \SI{115.29}{\micro\meter}.
And with it the elongation of the signal $x =\SI{230.58}{\micro\meter}$, which is consistent with  Fig.~\ref{fig2:Expvssim_2D}(a).

Upon laser excitation at fluences below the melting threshold, a strain wave will form and propagate along the crystal.
This wave will distort the crystal lattice until a certain depth related to the speed of the strain wave, the depth being presented with a red line in Fig.~\ref{EM1:Sketch_BorrmannFan}(c). 
For a delay $t = \SI{900}{\pico\second}$, the strain wave will have traveled at the  $\upsilon_\text{L}$ to a depth of \SI{7.6}{\micro\meter}, most of the crystal (\SI{292.4}{\micro\meter}) would not be distorted yet. 
There will be a small area of this crystal, as represented in the sketch with the white triangle, where the diffracted photons will almost not interact with the re-diffracted photons from the strain area.
If we calculate the non-distorted surface of the BF for this new crystal, we obtain $\delta$'$ = \SI{112.38}{\micro\meter}$.
This means that the affected BF is \SI{118.19}{\micro\meter}, which is consistent Fig.~\ref{fig2:Expvssim_2D}(b). 
In this way, using this trigonometry relation, we can locate the position in depth of the strain wave for the first instants after the laser excitation.


\subsection{\label{app:Thomsen model} Analytical Thomsen model}
\begin{figure}[b] 
\includegraphics[scale = 0.32]{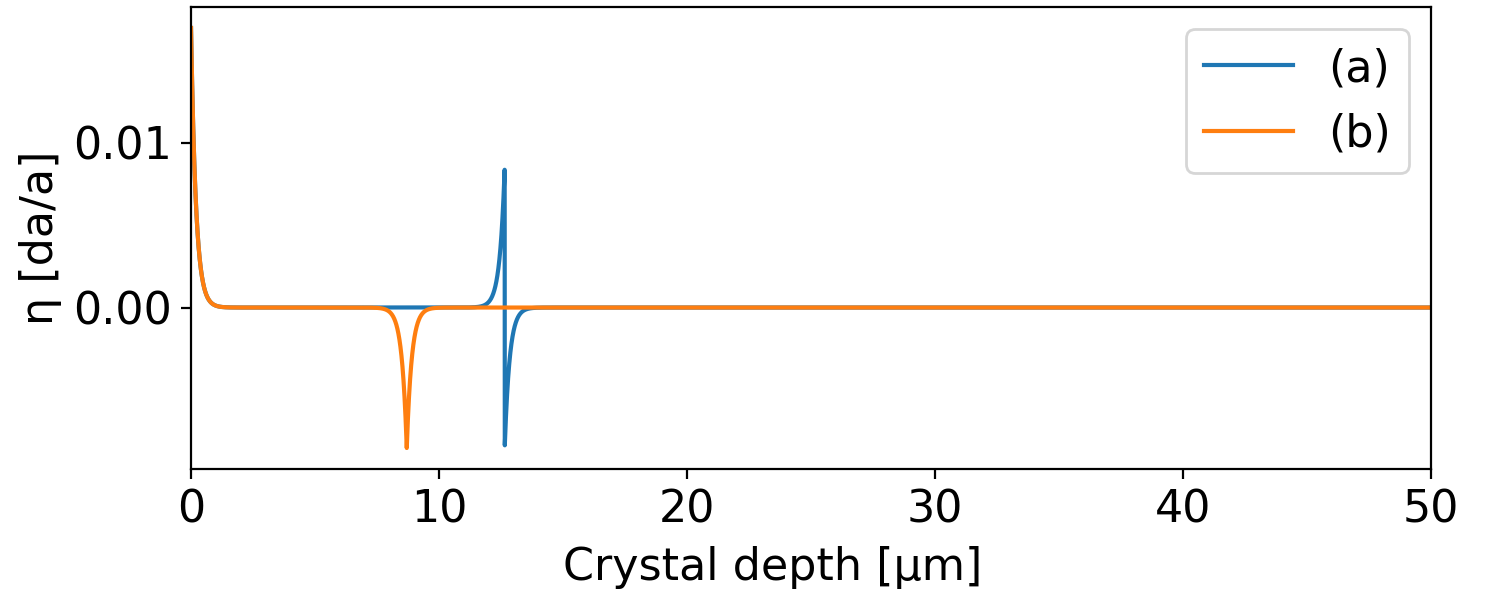}
\caption{\label{EM2:Strain_depth_profiles} Simulated strain profile for a \SI{300}{\micro\meter} Si crystal along the (a) longitudinal and (b) transversal or orthogonal direction to the surface of the crystal following the analytical solution presented in Eq.~\ref{eq.ThomsenPer} and Eq.~\eqref{eq.ThomsenPar}, respectively \cite{Thomsen96}.
}
\end{figure}

Considering the highly complex excitation and initial relaxation mechanisms with their corresponding large and interrelated parameter landscape, to date not quantitatively solved in the scientific community, our model needs to rely on simplifications. 
We based our model in the work of Thomsen et al., who

proposed an analytical 1D model to describe the generation of a stress pulse in a crystalline material after laser excitation \cite{Thomsen96}.
The model uses the optical, electronic and acoustic properties of the material to represent the lattice distortion at different time delays.
Following the model, the strain in the perpendicular direction to the surface $\eta_{33}$ in a material as a function of depth $z$ and delay time $t$ would be:

\begin{equation}
\eta_{33} = \eta_\text{A} \left[e^{-\frac{z}{\zeta}}\left(1-\frac{1}{2}e^{-\frac{\upsilon_\text{L} t}{\zeta}}\right)-\frac{1}{2} e^{-\frac{|z-\upsilon_\text{L} t|}{\zeta}}\text{sgn}(z-\upsilon_\text{L} t)\right]
\label{eq.ThomsenPer}
\end{equation}
with
\begin{equation}
\eta_\text{A} = (1-R) \frac{F \beta}{\zeta C} \frac{1+\nu}{1-\nu} \nonumber
\end{equation}

where $\upsilon_\text{L}$ is the longitudinal sound velocity, $R$ the reflectivity of the material, 
$F$ is the laser Fluence, $\beta$ the linear expansion coefficient, $C$ is the specific heat per unit volume, $\zeta$ is the absorption length and $\nu$ the Poisson ratio. 

Similarly, we can describe the lattice displacement parallel to the surface,  {$\eta_{hh0}$}, as presented in Eq.~\eqref{eq.ThomsenPar}.
In our model we have added an scaling factor, $I_{adj}$, that can be used to modulate the amplitude of the lattice distortion. 


\begin{equation}
\eta_{(hh0)} = I_\text{adj}\, \eta_\text{A} \left[e^{-\frac{z}{\zeta}}\left(1-\frac{1}{2}e^{-\frac{\upsilon_\text{T} t}{\zeta}}\right)-\frac{1}{2} e^{-\frac{|z-\upsilon_\text{T} t|}{\zeta}}\right]
\label{eq.ThomsenPar}
\end{equation}


Figure~\ref{EM2:Strain_depth_profiles} shows the simulated lattice distortion for a time delay of 900 ps produced by a \SI{50}{mJ / cm^2}, \SI{800}{\nano\meter} femtosecond laser pulse that excites the front surface of a Si crystal for both the perpendicular Fig.~\ref{EM2:Strain_depth_profiles}(a) and orthogonal Fig.~\ref{EM2:Strain_depth_profiles}(b) directions.
For it we have used the solution in Eq.~(\ref{eq.ThomsenPer}) and our proposed approximation for the orthogonal direction is presented in Eq.~(\ref{eq.ThomsenPar}).
Figure~S14 in the SM presents the waterfall plot of the strain propagation along the depth for the two contributions as used in the simulations.

This model simplifies the initial temperature profile to be exponential with an absorption depth, we have used this as a fitting parameter to obtain the best match with the experimental signal. 
A second important unknown parameter is the type of strain profile in this direction. 
Initial simulations performed with COMSOL \cite{Comsol} to replicate the material behavior under the laser excitation, showed a small uni-polar component (more than hundred times smaller than the longitudinal component) propagating into the depth at the transverse speed of sound,$\upsilon_\text{T}$, into the material depth as presented in Fig.~S15 \cite{SM20}.
We decided to keep the same uni-polar excitation for our wave based in these simulations, as we do not have certainty of the shape of the wave propagating into the depth. 
We believe that a truly quantitative comparison between experimental and model is not possible at this stage.
Yet, our model is able to reproduce the main features of the transient signal and therefore allows us to extract a certain amount of relevant information on the processes.

\subsection{\label{app:Simulations} Simulations and divergence}

\begin{figure}[b] 
\includegraphics[scale = 0.4]{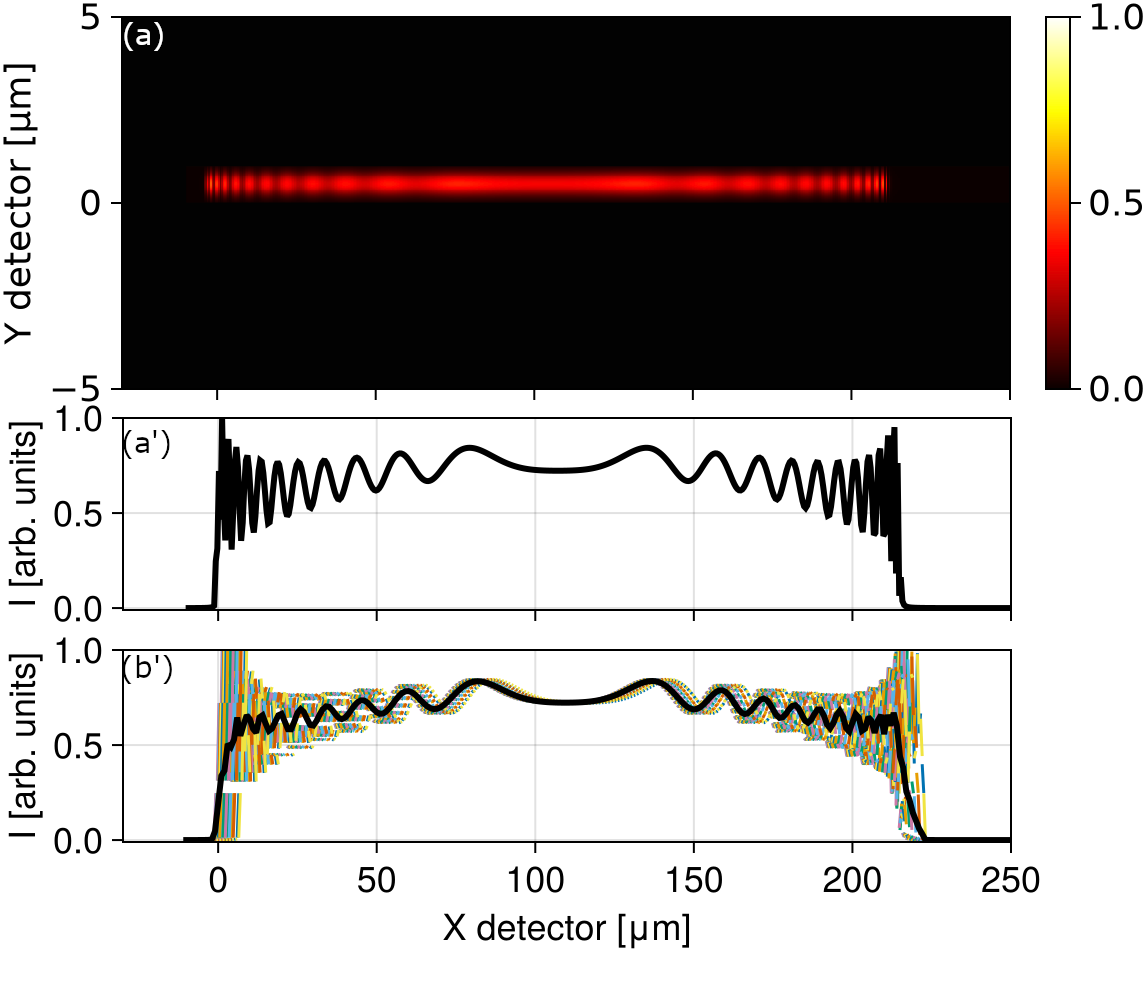}
\caption{\label{EM3:Simulation}
(a) Simulation of the diffraction wavefront for the $(220)$ Laue symmetric reflection using a single energy of \SI{9}{\kilo \electronvolt} for a \SI{300}{\micro\meter} thick Si crystal.
(a') Profile along the x detector direction of the simulated wavefront along the x direction of the detector for a single energy.
(b') (Colors) Profiles along the x detector direction of the simulation wavefronts using $21$ energies in the range \SI{8.990}{\kilo \electronvolt} and \SI{9.010}{\kilo \electronvolt} in steps of \SI{1}{\electronvolt}. (Black) Mean profile for the $21$ simulated wavefronts.
}
\end{figure}

The simulations were performed using an existing UDD code developed for the works presented in \cite{ARF18,ARF21,ARF23} rewritten to the Julia programming language \cite{Bezanson_Julia_A_fresh_2017}.
The parallelization in Julia has allowed to increase the performance by 
more than 20x per calculation layer.
The code takes into account parameters such as miscut of the crystal, energy of the x-ray, divergence of the x-ray beam or thickness,
which have been refined during the data simulation.
Figure~~\ref{EM3:Simulation}(a) presents the simulation of the diffracted wavefront for a single energy. 
Fig.~\ref{EM3:Simulation}(a') presents the profile of the simulated signal.
For a purely monochromatic experiment, it would be possible to distinguish all the different echoes maxima, obtaining more information about the effects in the layers close to the surface. 

In addition to the non-monochromaticity conditions in our work, the sample was located away from the focus in a location where the wavefront was not parallel, as presented in Fig. \ref{fig1:Sketch_MID}.
This curved wavefront together with possible inhomogeneities of the crystal sample at the surface are two possible explanations for the smearing out of the echoes at the edges.
To take into account this curvature, we have added to our simulations the effect of the divergence of the incident beam, by averaging a range of energies equivalent to \SI{20}{\electronvolt} around the main energy of the experiment \SI{9}{\kilo \electronvolt} with steps of \SI{1}{\electronvolt}.
The average of these $21$ energy steps reduces the fringes at the edges of the signal, leading to a better match with the experimental data as presented in Fig.~\ref{fig2:Expvssim_2D}(a$'$) and (b$'$).
For more detail, Fig.~\ref{EM3:Simulation} (b') presents the different simulated profiles for each of the 21 energies used in the simulation, together with the mean profile as presented in ~Fig.~\ref{fig2:Expvssim_2D}(a$'$).
A blurring of the mean signal profile (black line) is clearly observable at the edges.





%
%


\bibliography{MID_2024_UDD_SW}

\end{document}